\documentclass[12pt]{article}
\usepackage{bezier}

\newcommand{\levcon}[3]{\bigg\{{#1 \atop {#2#3}}\bigg\}}
\newcommand{\con}[3]{{\Gamma^#1}_{#2#3}}
\newcommand{\kon}[3]{{K^#1} _{#2#3}}
\newcommand{\tor}[3]{{T^#1}_{#2#3}}
\newcommand{\riem}[4]{{R^#1}_{#2#3#4}}
\newcommand{\tormid}[3]{{T_#1}{^#2}{_#3}}
\newcommand{\torup}[3]{T^{#1#2#3}}
\newcommand{\tordown}[3]{T_{#1#2#3}}
\begin{document}
\title{Torsion Degrees of Freedom in the Regge Calculus as
  Dislocations on the Simplicial Lattice} 

\author{J\"urgen
  Schmidt\thanks{Corresponding author, e-mail:jschmidt@agnld.uni-potsdam.de}\\Institut f\"ur Physik\\Lehrstuhl Nichtlineare Dynamik\\Universit\"at
  Potsdam\\D-14469 Potsdam, Germany\\[0.5cm]
  Christopher Kohler\\Institut f\"ur Theoretische und Angewandte
  Physik\\Universit\"at Stuttgart\\D-70550 Stuttgart, Germany\\[0.5cm]Proposed running title:\\Torsion in the Regge Calculus} 

\maketitle

\begin{abstract}
  Using the notion of a general conical defect, the Regge Calculus is
  generalized by allowing for dislocations on the simplicial lattice 
  in addition to the usual
  disclinations. Since disclinations and dislocations correspond to
  curvature and torsion singularities, respectively, the method we
  propose provides a natural way of discretizing
  gravitational theories with torsion degrees of freedom like the
  Einstein-Cartan theory. A discrete version of the Einstein-Cartan
  action is given and field equations are derived, demanding
  stationarity of the action with respect to the discrete variables of
  the theory.
\end{abstract}

\section{Introduction}
The theory we now call Regge Calculus was proposed in 1961 by T.\ 
Regge \cite{regge} as a discrete version of General Relativity
formulated within the framework of Riemannian geometry. In spite of
the experimental success of General Relativity, a number of authors
started to put gravitational theory on the grounds of non-Riemannian
geometry: The notion of Riemannian curvature was first generalized by
Cartan \cite{cartan}, introducing torsion degrees of freedom, and
later this concept was included in a formulation of gravitation as a
gauge theory of the Poincar\'{e} group by Sciama \cite{sciama} and
Kibble \cite{kibble} and worked out by Hehl et al.\ \cite{hehl1} (the
theory presented in the latter reference will be referred to as
Einstein-Cartan theory in the following). An even more general
geometry was proposed by Hehl et al.\ \cite{hehl2}, allowing for
nonmetricity in addition to curvature and torsion degrees of freedom.
An alternative approach to gravity are the so called
teleparallel theories \cite{hayashi}, based on the Weitzenb\"ock
geometry, working with torsion and vanishing curvature.

Despite this development in the continuum theory, only few attempts
have been made to include concepts of non--Riemannian geometry into
the Regge Calculus. Caselle et al.\ \cite{caselle} formulated Regge
Calculus as a lattice gauge theory of the Poincar\'{e} group and
pointed out the possibility of including torsion as closure failures
of the building blocks of the simplicial manifold (see also
\cite{gronwald}).  Drummond \cite{drummond} described torsion on the
d-dimensional Regge lattice as a piecewise constant tensor field,
i.e.\ within every d-simplex the torsion field was assigned a constant value,
which in general changes discontinuously at the hypersurface between two
neighboring simplices. Thus, in Drummond's approach the geometric
quantities curvature and torsion are treated in a different way: While
curvature appears on the lattice as a conical defect (a disclination)
of the underlying simplicial manifold, torsion does not correspond to
the simplicial structure itself, but is dealt with in a way similar to
Sorkin's treatment of the electromagnetic field on the simplicial
lattice \cite{sorkin}.

On the other hand the notion of torsion singularities, appearing as a
conical defect (a dislocation), has recently been discussed in the
literature again \cite{tod}. Its application to the theory of crystal
defects has been known for a long time \cite{kondo}, \cite{bilby},
\cite{kroner}, \cite{gairola}, \cite{holz2}, and the connection to
gravitation is pointed out in \cite{holz1}, \cite{katanaev},
\cite{kohler1}, and \cite{kohler2}. This suggests a natural way of
incorporating torsion degrees of freedom into the Regge Calculus much
in the same way as curvature: treating it as a conical defect of the
simplicial manifold.

In this paper, we will apply this idea in order to find a discrete
version of the Einstein-Cartan theory, i.e.\ we will construct the
discrete analogue of the Einstein-Cartan action, choose appropriate
sets of discrete variables, and compute the corresponding field
equations.

\section{Simplicial Torsion}
In the presence of torsion, infinitesimal parallelograms in space-time
generally do not close, i.e.\ to a surface element $dx^\mu\wedge
dx^\nu$ there belongs a closure failure $dG^\alpha$ proportional to
the torsion tensor
\begin{equation}
\label{three}
dG^\alpha = \tor{\alpha}{\mu}{\nu}\,dx^\mu\wedge dx^\nu\,.
\end{equation}
Here, greek indices take on the values $0,\ldots,\mbox{d}-1$ where d
is the dimension of space-time.

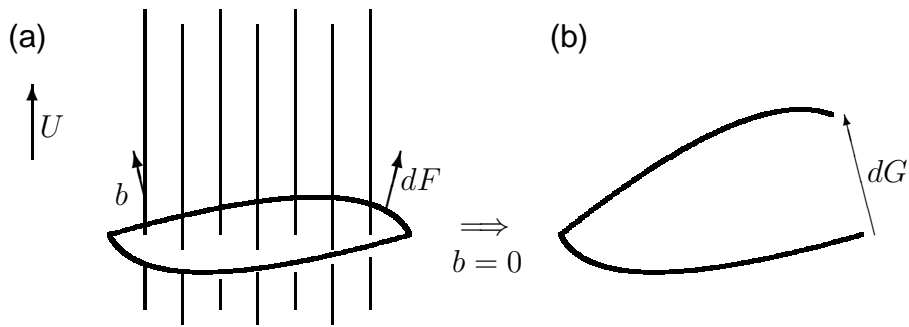
\begin{figure}[ht]
\begin{picture}(125,50)
\linethickness{0,5mm}
\put(0,-15){
\begin{picture}(70,65)

\bezier{200}(20,30)(55,40)(60,30)
\bezier{200}(20,30)(25,20)(60,30)
\thicklines

\multiput(25,30)(10,0){4}{\line(0,1){30}}
\multiput(30,28)(10,0){3}{\line(0,1){30}}
\put(25,20){\line(0,1){6}}
\put(35,20){\line(0,1){4}}
\put(45,20){\line(0,1){5}}
\put(55,20){\line(0,1){7}}
\put(30,18){\line(0,1){7}}
\put(40,18){\line(0,1){7}}
\put(50,18){\line(0,1){8}}
\put(10,40){\vector(0,1){10}}
\put(11,43){$ U$}
\put(57,33){\vector(1,4){2}}
\put(59,36){${dF}$}
\put(25,35){\vector(-1,4){1.5}}
\put(21,34){$ b$}
\end{picture}}
\put(68,15){$\Longrightarrow$}
\put(67,10){$ b=0$}
\put(8,40){(a)}
\put(80,40){(b)}
\put(80,-15){
\begin{picture}(45,65)

\bezier{200}(0,30)(25,50)(36,46)
\bezier{200}(0,30)(5,20)(40,30)
\put(42,30){\vector(-1,4){4}}
\put(41,37){${dG}$}

\end{picture}}
\end{picture}
\caption[Der simpliziale Torsionstensor]
{\label{simptor}\it (a) Loop around a bundle of dislocations and
  (b) the same loop in defect free medium.}
\end{figure}

We assume that the tensor of the torsion density is of the form
\begin{equation}
\tor{\alpha}{\mu}{\nu}\sim b^\alpha S_{\mu\nu}
\end{equation}
where the Burgers vector $b^{\alpha}$ gives the strength and direction of the
associated closure failure and the antisymmetric tensor $S_{\mu\nu}$ the orientation of the support of the distributional torsion
field. This can be motivated in a heuristic way similar to Regge's
\cite{regge} construction of his simplicial curvature tensor: We take a
bundle of parallel dislocations in three dimensions (e.g.\ in an
elastic medium, generated by a Volterra Process \cite{puntigam}) each
with the same Burgers vector $b^{\alpha}$ (see Fig.\ \ref{simptor}), their
orientation \unitlength1mm
given by the unit vector ${U^{\alpha}}$. Now we encircle the bundle by a small
loop with normal vector ${dF^{\alpha}}$. Transferred to defect free space, this
loop does not close, the closure failure ${dG^{\alpha}}$ being proportional to
the Burgers vector ${b^{\alpha}}$. If ${dN}$ denotes the number of
dislocations enclosed by the loop, we have
\begin{eqnarray}
dG^\alpha &=& dN\,b^\alpha \nonumber\\
&=& \rho \,U^\lambda dF_\lambda \, b^\alpha\,,
\end{eqnarray}
where $\rho$ is the density of dislocations in a surface
perpendicular to the bundle. Inserting the dual quantities
\begin{eqnarray}
U^\lambda&=&\frac{1}{2}\epsilon^{\lambda\mu\nu}S_
{\mu\nu}\nonumber\\ dF_\lambda&=&\frac{1}{2}\epsilon_{\lambda\beta\gamma}
dF^{\beta\gamma}\,
\end{eqnarray}
leads to
\begin{equation} 
dG^\alpha=\frac{1}{2}\rho \,b^\alpha S_{\mu\nu}\,dF^{\mu\nu}\,.
\end{equation}
Comparison with equation (\ref{three}) yields
\begin{equation}
\label{five}
\tor{\alpha}{\mu}{\nu}=\frac{1}{2}\rho\,b^\alpha S_{\mu\nu}\,.
\end{equation}
In d dimensions, the defect has codimension two, i.e.\ its orientation
is given by d--2 orthogonal unit vectors ${U^{\alpha}}_1,\ldots ,{U^{\alpha}}_{d-2}$. The
antisymmetric tensor $S_{\mu\nu}$ is then defined by
\begin{equation}
S_{\mu\nu}=\epsilon_{\mu\nu\alpha_1\ldots\alpha_{d-2}}U_1^{\alpha_1}\ldots
U_{d-2}^{\alpha_{d-2}}\,.
\end{equation}

\section{Discrete Action, Dynamical Variables, and Field Equations} 
In the geometry of a continuous manifold, torsion enters the scene when
we allow the connection $\con{\alpha}{\mu}{\nu}$ to be nonsymmetric,
its antisymmetric part defining the torsion tensor
\begin{equation}
\tor{\alpha}{\mu}{\nu}\equiv\con{\alpha}{\mu}{\nu}-\con{\alpha}{\nu}{\mu}\,.
\end{equation}
Assuming the condition of metricity $g_{\mu\nu ;\lambda}=0$, the connection reads
\begin{equation}
\label{one}
\con{\alpha}{\mu}{\nu}=\levcon{\alpha}{\mu}{\nu}+\kon{\alpha}{\mu}{\nu}
\end{equation}
where
\begin{equation}
\kon{\alpha}{\mu}{\nu}\equiv\frac{1}{2}(\tor{\alpha}{\mu}{\nu}-\tormid
{\mu}{\alpha}{\nu}-\tormid{\nu}{\alpha}{\mu})
\end{equation}
defines the contortion tensor and 
\begin{equation}
\levcon{\alpha}{\mu}{\nu}=\frac{1}{2}g^{\alpha\rho}(\partial_\mu g_{\nu\rho}
+\partial_\nu g_{\mu\rho}-\partial_\rho g_{\mu\nu})
\end{equation}
is the Levi-Civita connection of General Relativity. In terms of the
connection $\Gamma^{\alpha}_{\mu\nu}$, the curvature tensor is defined as
\begin{equation}
\riem{\alpha}{\beta}{\mu}{\nu}\equiv\partial_\mu\con{\alpha}{\beta}{\nu}
-\partial_\nu\con{\alpha}{\beta}{\mu}+\con{\alpha}{\gamma}{\mu}
\con{\gamma}{\beta}{\nu}-\con{\alpha}{\gamma}{\nu}\con{\gamma}{\beta}{\mu}\,.
\end{equation}
Using equation (\ref{one}) we can now split the curvature scalar into
a part depending only on $g_{\mu\nu}$ and on its first derivatives and
a part depending only on the torsion. The result is
\begin{eqnarray}
\label{four}
R\equiv{R^{\alpha\beta}}_{\alpha\beta}&=&
g^{\alpha\beta}\left(\levcon{\mu}{\alpha}{\nu}\levcon{\nu}{\beta}{\mu}-
\levcon{\mu}{\alpha}{\beta}\levcon{\nu}{\mu}{\nu}\right)\nonumber\\
& &
+\frac{1}{4}\torup{\alpha}{\mu}{\nu}\tordown{\alpha}{\mu}{\nu}
+\frac{1}{2}\torup{\alpha}{\mu}{\nu}\tordown{\nu}{\mu}{\alpha}
+\tor{\alpha}{\alpha}{\mu}{T^{\nu\mu}}_\nu \nonumber\\
& &+\frac{1}{\sqrt{-g}}\,\mbox{divergence}\,,
\end{eqnarray}
where the last term does not contribute to the action integral when we
compute it for a manifold without boundary.

With the splitting (\ref{four}) of the curvature scalar, the
Einstein-Cartan action reads
\begin{eqnarray}
\label{two}
S_{EC}&\equiv&\frac{1}{2}\int d^dx\sqrt{-g}R(g,T)\nonumber\\
      &=&S_{EH}
+\frac{1}{2}\int d^dx\sqrt{-g}\left(
\frac{1}{4}\torup{\alpha}{\mu}{\nu}\tordown{\alpha}{\mu}{\nu}
+\frac{1}{2}\torup{\alpha}{\mu}{\nu}\tordown{\nu}{\mu}{\alpha}
+\tor{\alpha}{\alpha}{\mu}{T^{\nu\mu}}_\nu\right)\, \nonumber\\
\end{eqnarray}
where we have identified the Einstein--Hilbert action
\begin{equation}
S_{EH}=\frac{1}{2}\int d^dx\sqrt{-g}
g^{\alpha\beta}\left(\levcon{\mu}{\alpha}{\nu}\levcon{\nu}{\beta}{\mu}-
\levcon{\mu}{\alpha}{\beta}\levcon{\nu}{\mu}{\nu}\right)\,.
\end{equation}
Discretization of the Einstein-Hilbert action leads to the Regge
action \cite{regge}, so we are concerned here with the second term in
equation (\ref{two}).

In Regge Calculus, the curvature scalar is defined as a
distribution with support on the d--2 dimensional hypersurfaces of the
lattice (in Regge Calculus called the bones of the lattice). In the
same way we treat the non-Riemannian curvature scalar, in particular its non-Riemannian part,
i.e.\ we define {\it squares} of the torsion tensor as a
distribution. Using equation (\ref{five}) we obtain for the
contribution of the $i$-th bone to a typical term appearing under the
Einstein-Cartan action integral (eq.\ (\ref{two})) [no sum over repeated latin indices]
\begin{equation}
T^{\alpha\mu\nu}_{(i)} T_{{(i)}\alpha\mu\nu}
=\frac{1}{4}b^\alpha_{(i)} b_{(i)\alpha} S_{(i)}^{\mu\nu}S_{{(i)}\mu\nu}
\int ds_{(i)}^1 \dots ds_{(i)}^{d-2}\delta^{\left(\mbox{d}\right)} \left\{{ x}-{ y}_{(i)}
(s_{(i)}^1,\dots,s_{(i)}^{d-2})\right\}\,.\hspace{1.0cm}
\end{equation} 
Here, ${y_{(i)}}(s^1_{(i)},\ldots,s^{d-2}_{(i)})$ is the set of points of the
$i$-th bone, parameterized by $s^1_{(i)},\ldots,s^{d-2}_{(i)}$. With
\begin{equation}
S_{\mu\nu}S^{\mu\nu}=2,\quad
S^{\mu\nu}S_{\mu\alpha}=\delta_\alpha^\nu-\left({U_1}_\alpha{U_1}^\nu+\ldots
+U_{(d-2)\alpha}{U_{d-2}}^\nu\right)\,,
\end{equation}
we obtain
\begin{eqnarray}
\nonumber
& \frac{1}{4}T_{(i)}^{\alpha\mu\nu}T_{{(i)}\alpha\mu\nu}+\frac{1}{2}T_{(i)}^{\alpha
\mu\nu}T_{{(i)}\nu\mu\alpha}+{T_{(i)}^\alpha}_{\alpha\mu}{T_{(i)}^{\nu\mu}}_\nu
\hspace{5cm} &
\\
\label{six}
& \hspace{2cm} =\frac{1}{8}b_{(i)}^2\int ds_{(i)}^1 \dots ds_{(i)}^{d-2}\delta^{\left(\mbox{d}\right)} 
\left\{{ x}-{ y}_{(i)}(s_{(i)}^1,\dots,s_{(i)}^{d-2})\right\}\,, &
\end{eqnarray}
with
\begin{equation}
\label{eight}
b_{(i)}^2\equiv\left((b_{{(i)}\alpha} U_{(i)1}^\alpha)^2+\ldots+(b_{{(i)}\alpha}
U_{(i){d-2}}^\alpha)^2\right)\,,
\end{equation}
where $b_{(i)}$ denotes the part of the Burgers vector that is {\it
  parallel} to the bone, i.e.\ only screw dislocations contribute to
the action.

Summing equation (\ref{six}) over all the bones of the lattice and
integrating over the whole manifold yields
\begin{displaymath}
\int dV\left(
\frac{1}{4}\torup{\alpha}{\mu}{\nu}\tordown{\alpha}{\mu}{\nu}
+\frac{1}{2}\torup{\alpha}{\mu}{\nu}\tordown{\nu}{\mu}{\alpha}
+\tor{\alpha}{\alpha}{\mu}{T^{\nu\mu}}_\nu\right)\nonumber
\end{displaymath}
\begin{eqnarray}
&=&\frac{1}{8}\sum_i b_{(i)}^2 \int dV 
\int ds_{(i)}^1 \dots ds_{(i)}^{d-2}\delta^{\left(\mbox{d}\right)} 
\left\{{ x}-{ y}_{(i)}(s_{(i)}^1,\dots,s_{(i)}^{d-2})\right\}\nonumber\\
&=&\frac{1}{8}\sum_ib_{(i)}^2A_{(i)}\,,
\end{eqnarray}
where $A_{(i)}$ is the (d--2 dimensional) area of the $i$-th bone. This is
the simplicial analogue to the second term in (\ref{two}). For the
Einstein-Hilbert term we substitute the Regge action and obtain the
lattice action
\begin{equation}
\label{reggecartanw}
S=\sum_i\left(\varphi_{(i)}+\frac{1}{16}b^2_{(i)}\right)A_{(i)}
\,,
\end{equation}
where the $\varphi_{(i)}$ are the deficit angles of the bones.

As the first set of dynamical variables of the theory, we take the
link lengths $l_{(i)}$ of the lattice (corresponding to the components of
the metric tensor of the continuum theory) as in the usual Regge
Calculus. In the Einstein-Cartan theory, the components of the torsion
tensor itself become the second set of variables of the action, so the
Burgers vectors $b_{(i)}$ of the bones are the most appropriate choice
(see eq.\ (\ref{five})) of lattice variables in a discretized Einstein-Cartan theory. Since only the projections of
the Burgers vectors onto the bones contribute to the action, we
choose the $b_{(i)}$ defined in (\ref{eight}) as (scalar) variables
representing torsion on the simplicial lattice.

Adding a matter term $L(\{l_{(j)}\},\{b_{(j)}\})$ to the action, variation
with respect to $l_{(i)}$ and $b_{(i)}$ gives two sets of field equations,
\begin{equation}
\label{seven}
\sum_i \left( \varphi_{(i)} +\frac{1}{16} b_{(i)}^2 \right) \frac{\partial A_{(i)}}
{\partial l_{(j)}}=-\frac{\partial L}{\partial l_{(j)}} 
\end{equation}
and
\begin{equation}
b_{(j)}=-\frac{1}{A_{(j)}}\frac{\partial L}{\partial b_{(j)}} \,.
\end{equation}
In four dimensions we have
\begin{equation}
\frac{\partial A_{(i)}}{\partial l_{(j)}}=\cot\theta_{(ij)}\,,
\end{equation}
where $\theta_{(ij)}$ is in the $i$-th bone (triangle of the lattice)
the angle opposite to the $j$-th link. So, the first field equation
reads
\begin{equation}
\sum_i \left( \varphi_{(i)} +\frac{1}{16} b_{(i)}^2 \right)\cot\theta_{(ij)}=
-\frac{\partial L}{\partial l_{(j)}}\,, 
\end{equation}
where the sum now extends over all the bones having the link $l_{(j)}$ in
common.

The Burgers vector couples algebraically to the matter term $\partial
L/\partial b_{(j)}$, i.e.\ we have no dislocations on the lattice
 in vacuum. Without
this matter term, which is interpreted in continuum theory as the spin
density of matter, the field equations reduce to the ordinary Regge
equations. This is in parallel to the result of the continuum theory,
where the Einstein-Cartan field equations reduce to the Einstein
equation in vacuum.

\section{Conclusion}
From the application of non-Riemannian geometry to the theory of
defects in solids \cite{kondo},\cite{bilby}, we know that the torsion
field can be interpreted as a continuous distribution of dislocations.
Vice versa, we can express the torsion induced by a single dislocation
as a delta like distribution with support on the dislocation line.
This suggests the idea of regarding torsion degrees of freedom within
the Regge Calculus as dislocations of the lattice, the bones carrying
singular torsion in addition to a curvature singularity. Thus, to a
loop around a bone, we assign not only a rotation (of a test vector),
but also a translation, which we call the Burgers vector of the bone.
Discretizing the Einstein-Cartan action, we find a generalization of
the Regge action where the Burgers vector of a particular bone leads
to a shift of the deficit angle. Variation of the action with respect
to the link lengths and the Burgers vectors leads to two sets of field
equations that show fundamental properties of the continuum
equations.

Fermion fields, defined on the lattice, would lead to a nonvanishing
spin density and act as a source of torsion \cite{ren} localized at
the matter. Although classically in vacuum not possible, in simplicial
quantum gravity configurations with nonvanishing Burgers vectors could
be important as quantum fluctuations of the lattice
\cite{caselle}.\vspace{0.5cm}

\noindent
{\em Acknowledgment}\\
We have benefited from discussions with A.\ Holz and T.\ Filk.

\end{document}